\documentclass{jfm}
\usepackage{epstopdf, epsfig}
\usepackage{siunitx}
\usepackage{url}
\usepackage{xcolor}
\usepackage{bm}
\usepackage{newtxtext}
\usepackage[varg]{newtxmath}

\shorttitle{Normal stress differences in dense suspensions}
\shortauthor{R. Seto, G. G. Giusteri}
\title{Normal stress differences in dense suspensions}
\author{Ryohei Seto\aff{1}
  \corresp{\email{setoryohei@me.com}},
  Giulio G. Giusteri\aff{2}
}

\affiliation{\aff{1} Department of Chemical Engineering, Kyoto University, Nishikyo-ku, Kyoto, 615-8510 Japan
\aff{2}Department of Mathematics, Politecnico di Milano, Piazza Leonardo da Vinci 32, 20133 Milano, Italy
}

\DeclareMathOperator{\tr}{Tr}
\newcommand*\Tensor[1]{\mathsfbi{#1}}
\newcommand{\cddot}{\mathbin{:}}
\newcommand{\secref}[1]{\S\ref{#1}}

\newcommand{\figref}[1]{\figurename~\ref{#1}}

\newcommand{\CauchyStress}{\bm{\sigma}}

\usepackage{amsmath}	
\begin{document}
\maketitle

\begin{abstract}
The presence and the microscopic origin of normal stress differences in dense suspensions under simple shear flows are investigated by means of inertialess particle dynamics simulations, taking into account hydrodynamic lubrication and frictional contact forces. The synergic action of hydrodynamic and contact forces between the suspended particles is found to be the origin of negative contributions to the first normal stress difference $N_1$, whereas positive values of $N_1$ observed at higher volume fractions near jamming are due to effects that cannot be accounted for in the hard-sphere limit. Furthermore, we found that the stress anisotropy induced by the planarity of the simple shear flow vanishes as the volume fraction approaches the jamming point for frictionless particles, while it remains finite for the case of frictional particles.
\end{abstract}

\section{Introduction}

Steady-shear rheology
provides a fundamental framework for the investigation and description of the properties of incompressible non-Newtonian fluids.
In the presence of a simple shear flow, with shear rate $\dot{\gamma}$, the response of different fluids is characterized by three degrees of freedom of the symmetric stress tensor $\CauchyStress$ (since the others are fixed by the geometry of the flow).
These are commonly identified with the shear stress $\sigma \equiv \langle \sigma_{xy} \rangle $, 
through which the viscosity $\eta \equiv \sigma/\dot{\gamma}$ is defined, and the first and second normal stress differences, 
$N_1 \equiv \langle \sigma_{xx}-\sigma_{yy} \rangle $ and $N_2 \equiv \langle \sigma_{yy}-\sigma_{zz} \rangle$, respectively. 
(We set $x$ as the flow direction, $y$ as the gradient direction, and $z$ as the vorticity direction
of the simple shear flow.)
Newtonian fluids are characterized by the constant value of $\eta$, while $N_1$ and $N_2$ are zero.
On the other hand, all of the three functions are required to identify or distinguish non-Newtonian fluids.


Historically, normal stress differences have been particularly important to characterize viscoelastic fluids.
In steady shear, it is impossible to distinguish the viscous and the elastic contribution to the shear stress, but the presence of a nonvanishing $N_1$ is a signature of elastic effects.
Indeed, the extensional component
of the flow stretches elastic elements, such as polymer chains, 
and the convection determined by the rotational component 
of the flow leads to positive values of $N_1$.
It is desirable to be able, also for other fluids, to associate nonvanishing values of the normal stress differences with some microscopic mechanism 
causing non-Newtonian behaviors.
%


Suspensions, namely mixtures of solid particles and viscous liquids,
are an important class of non-Newtonian fluids
that exhibit shear thinning and thickening \citep{Laun_1984, Barnes_1989, Mewis_2011, Guy_2015}.
Since suspended particles are usually very rigid,
there is no obvious elastic component in such fluids.
To thoroughly characterize them,
significant efforts have been made to measure normal stress differences~%
\citep{Laun_1994,Zarraga_2000,Singh_2003,Lootens_2005,Couturier_2011,Dai_2013,Dbouk_2013,Cwalina_2014,Royer_2016,Gamonpilas_2016,Pan_2017, Hsiao_2017}.
Though experimental results do not always agree,
most of them reported a negative $N_1$ for moderately 
dense suspensions at high shear rates.
Few of them also reported a characteristic transition 
from negative to positive values of $N_1$ at high shear rates in very dense suspensions.
By contrast with the positive $N_1$ measured for viscoelastic fluids,
negative values of $N_1$ are considered an unusual
and unique rheological feature of suspensions.
%


The question of what microscopic effects determine the observed normal stress 
differences in suspensions has so far received 
only partial answers (for more details see the recent review article by \citet{Guazzelli_2018}).
Stokesian Dynamics simulations by \citet{Phung_1996} reproduced negative values of $N_1$ 
at relatively high shear rates, meaning large values of the P\'eclet number $\Pen$.
\citet{Bergenholtz_2002} presented a theoretical argument identifying
a negative hydrodynamic contribution to $N_1$ and a positive one due to Brownian interactions in dilute suspensions.
More recent simulations, including frictional contacts besides hydrodynamic interactions,
reproduced the transition 
from negative to positive $N_1$~\citep{Mari_2015a,Singh_2018,Boromand_2018}.
As a consequence, positive values of $N_1$ in very dense suspensions tend to be explained 
as an effect of frictional contact forces or granular dilatancy.
%


This fostered the misconception that the sign of $N_1$ can discriminate between regimes in which either hydrodynamic interactions are dominant (negative $N_1$) or contact interactions are (positive $N_1$).
With the present paper, 
we provide evidence that a different interpretation is in order. 
Namely, upon increasing the volume fraction $\phi$ in the high-P\'eclet-number limit,
there is a transition from a regime in which the negative values of $N_1$ 
are essentially determined by hydrodynamic interactions 
to a regime in which synergies between hydrodynamic and contact interactions produce even more evident negative values of $N_1$.
It is only at volume fractions approaching the jamming conditions that we can observe positive values of $N_1$ and our results indicate that, for a computational model that aims at simulating hard-sphere suspensions, these ought to be regarded as artifacts of the numerical approximation.
Indeed, we can identify the origin of a positive $N_1$ in the elastic interactions employed to regularize the hard-sphere contacts.
In turn, this fact suggests that experimental measurements of positive values 
of $N_1$ may indicate the presence of elastic interactions,
such as soft elastic layers at particle surfaces or some cohesive bonding between particles,
that cannot be captured by simple hard-sphere models.
Another possible explanation for these observations traces them back to boundary effects,
due to presence of walls that cannot be avoided in standard rheometry~\citep{Yeo_2010,Gallier_2016}.
If this is the case, simulations of the bulk rheology, like the ones we performed, should not be expected to reproduce the measured values of $N_1$ near jamming.


A key point in our investigation is the geometric interpretation of $N_1$ as a proxy for the misalignment between the stress $\CauchyStress$ and the symmetric part of the velocity gradient $\Tensor{D}$~\citep{Giusteri_2018}.
Indeed, the ratio $N_1/\sigma$ determines the angle $\theta_{\mathrm{s}}$, 
in the flow plane, between the eigenvectors of $\CauchyStress$ and those of $\Tensor{D}$.
Such misalignment does not occur in planar extensional flows~\citep{Seto_2017}.
Another fruitful heuristic step involves the approximation of $N_1$ 
with the first normal stress difference generated only by normal forces between pairs of particles. 
This allows us to draw a direct and pictorial link between the microstructure of the force network generated under simple shear and the macroscopic value of $N_1$,
which opens the way for extending a similar analysis to the case of granular flows.


To complete our analysis of normal stresses, we study the quantity $N_0 \equiv N_2 +N_1/2$, 
that measures a stress anisotropy caused by the planarity of simple shear flows.
Differently from the standard $N_2$, the quantity $N_0$ is fully independent of $N_1$,
since they relate to mutually orthogonal terms 
in a linear decomposition of the stress tensor~\citep{Giusteri_2018}.
In this sense, $N_0$ is more informative than $N_2$,
as it appears also from its use 
in presenting experimental measurements~\citep[see, for instance,][]{Boyer_2011a}.

\section{Computational model}

The rheology of dense suspensions is dominated by the shear-induced microstructure but, 
except for a few asymptotic regimes, a theoretical treatment of the problem is out of reach.
We employ a simulation model developed in previous works~\citep{Seto_2013a,Mari_2014},
aiming to reproduce inertialess particle dynamics in Stokes flows.
Our simulation is similar to Stokesian Dynamics~\citep{Brady_1988},
but it omits long-range hydrodynamic interactions as explained below.
%


It is known that
the original Stokesian Dynamics simulation
is singular in the non-Brownian limit ($\Pen \to \infty$)
due to a diverging factor of $1/h$ in the lubrication resistance~\citep{Ball_1995},
where $h$ is the interparticle gap.
Since this singularity is due to the mathematical idealization, we can obtain some physical insight by using a slightly modified model.
We thus regularize the lubrication resistance
by introducing a small length scale $\delta$ and replacing the factor $1/h$ with $1/(h+\delta)$.
Though this is a reasonable modification, it has a drastic consequence: particle contacts are no longer forbidden.
We then need to introduce also a contact model.
%

Particles in suspensions are very hard, 
so that deformations under typical stresses are negligibly small
and we can consider them as rigid bodies.
A simulation strategy for the dynamics of 
hard spheres with multiple contacts,
in which overlaps are perfectly avoided, is available~\citep{Lerner_2012a}.
However, this approach is only for frictionless systems,
where particles can freely slide against each other.
In real systems, particles may also experience some tangential contact forces.
To be able to capture these effects,
we employ a frictional contact model 
commonly used in Discrete Element Methods.
At each contact point, normal and tangential forces are activated.
The strength $|\bm{F}_{\mathrm{C}}^{n} |$ of the normal repulsive force  
is proportional to the overlap $|h|$ between two particles,
$|\bm{F}_{\mathrm{C}}^{n} | =  k_{\mathrm{n}} |h|$.
Though the constant $k_{\mathrm{n}}$ could match the real elastic modulus of the particles,
we usually need to set a smaller value to capture the contact dynamics with reasonably large time steps
(more on this point in \secref{123947_14Jun18}). 
The strength $|\bm{F}_{\mathrm{C}}^{t}|$ of the tangential force 
is proportional to the sliding displacement at the contact point,
and the proportionality constant $k_{\mathrm{t}}$ 
is set to be half of $k_{\mathrm{n}}$ in this work.
Regarding the maximum tangential force, 
we implement a simple Coulomb friction model, 
where the static friction coefficient $\mu$ determines 
the bound $|\bm{F}_{\mathrm{C}}^{t} | < \mu |\bm{F}_{\mathrm{C}}^{n} |$ 
\citep[see ][ for details]{Mari_2014}. 


Thanks to the regularization of the lubrication resistance
and the introduction of an effective contact model 
we can simulate arbitrarily high values of $\Pen$ 
and we focus our attention on the infinite-$\Pen$ limit.
This regime could not be explored by previous theoretical studies developed 
in the low-$\Pen$ regime~\citep{Brady_1995,Bergenholtz_2002}.
In the infinite-$\Pen$ limit, we can neglect Brownian forces.
Then, most of the particles come into contact 
or in near-contact with others under shear flows.
In this case, for the investigated range of volume fractions 
the long-range hydrodynamic interactions
are much less important than the short-range lubrication interactions 
and we can thus ignore the former in our simulation.

We target sufficiently small particles in a viscous liquid,
whereby particle and fluid inertia 
do not play a role:
both the Stokes number and the Reynolds number 
are assumed to be zero.
In this Stokesian regime,
the particles obey the overdamped equations of motion in the form of a balance
\begin{equation}
\bm{F}_{\mathrm{H}} 
+
\bm{F}_{\mathrm{C}} 
= \bm{0}\label{162435_6Jun18}
\end{equation}
between hydrodynamic and contact forces, $\bm{F}_{\mathrm{H}}$ and $\bm{F}_{\mathrm{C}}$, respectively.
The hydrodynamic interactions (force and torque)
can be expressed as
the sum of linear resistances to the relative velocities and imposed deformation. 
We have
\begin{equation}
  \bm{F}_{\mathrm{H}} = 
- \Tensor{R}\cdot (\bm{U}-\bm{u})
+ \Tensor{R}'\cddot \Tensor{D},\label{162430_6Jun18}
\end{equation}
where $\Tensor{R}$ and $\Tensor{R}'$ are the resistance matrices.
The linear and angular velocities globally represented by the vector $\bm{U}$
can be determined by solving 
\eqref{162435_6Jun18} and \eqref{162430_6Jun18},
and particles are 
moved and rotated accordingly.
The simulation box of volume $V$ (on which periodic boundary conditions are imposed)
is deformed by following the simple shear flow $\bm{u} = \dot{\gamma} y \bm{e}_x$.
The stress tensor is given by
\begin{equation}
  \CauchyStress = 
  V^{-1} 
  \sum_i \Tensor{S}_{\mathrm{H}}^{(i)}
  +
  V^{-1} 
  \sum_{i >j } (\bm{r}^{(j)}-\bm{r}^{(i)}) \bm{F}_{\mathrm{C}}^{(ij)},
\end{equation}
where $\Tensor{S}_{\mathrm{H}}^{(i)}$ is the stresslet on the $i$-th particle, 
generated by hydrodynamic interactions.

We use an adaptive time step $\Delta t$ to update the particle positions 
based on the determined velocities
in such a way that a given maximum displacement $d_{\mathrm{max}}$
of the particles is respected in each step:
$\Delta t = d_{\mathrm{max}} / \max |\bm{U}^{(i)}|$.
The parameter $d_{\mathrm{max}}$ must be selected appropriately, depending on the value of the stiffness $k_{\mathrm{n}}$.
This procedure, necessary to avoid flaws such as inactive tangential contact force 
due to unphysical jumps,
makes the simulations for stiff particles near the jamming point very time-consuming.


\section{Results and discussion}
\begin{figure}
\centering
  \includegraphics[width=0.4\textwidth]{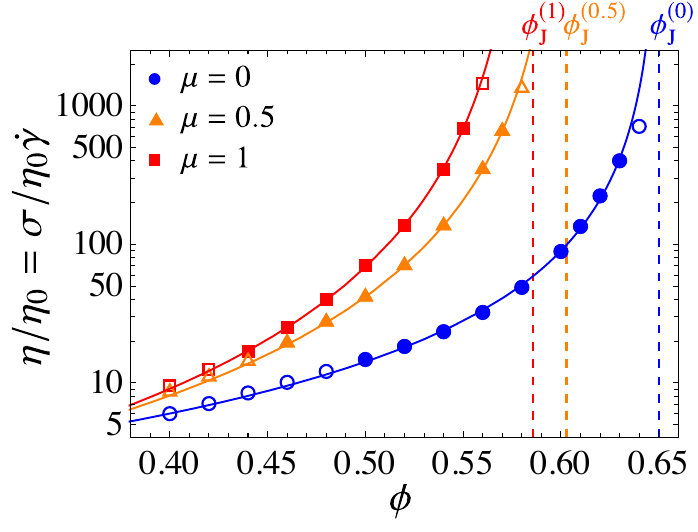}
  \caption{%
    The relative viscosity $\eta/\eta_0$ diverges as the volume fraction $\phi$ 
    approaches the jamming point $\phi_{\mathrm{J}}^{(\mu)}$, 
    a decreasing function of the friction coefficient $\mu$.
    Solid lines are obtained by fitting 
    $\eta/\eta_0 = c (\phi_{\mathrm{J}} - \phi)^{-\lambda}$
    to the data with the filled symbols.
    Vertical dashed lines represent
    the values of $\phi_{\mathrm{J}}^{(\mu)}$ estimated
    from the fitting for each $\mu$.
    The data corresponding to the open symbols near jamming are omitted
    in the fitting because of the potential inaccuracy due to the particle softness.
  }
  \label{163218_21May18}
\end{figure}

We work in the infinite-$\Pen$ limit and we are interested in exploring the dependence of the rheology of dense suspensions 
on the volume fraction $\phi$ of solid particles dispersed in a Newtonian fluid with viscosity $\eta_0$. 
The reported data are, unless specified otherwise, 
ensemble averages of time averages (taken in a statistically steady state) 
over 20 independent 3D simulations of 1000 particles, 
with a bidisperse size distribution characterized by size ratio $1.4$ 
and volume ratio about~$1$.
The cutoff length employed to regularize the lubrication singularity
is set to $\delta = 10^{-3}a$ \citep{Wilson_2002},
with $a$ being the radius of the smaller particles.
The parameter $k_{\mathrm{n}}$ for the contact model 
is set to $10^{5}k_0$, where the constant $k_0 \equiv 6 \pi \eta_0 a \dot{\gamma}$ 
is a reference value determined by the Stokes drag 
under a constant shear rate $\dot{\gamma}$.
(Our simulation is analogous to rate-controlled rheological measurements,
thus the shear rate $\dot{\gamma}$ is constant over time.)
The maximum particle displacement 
is set to $d_{\mathrm{max}}= 5 \times 10^{-4}a$ and the time 
step adaptively computed as described above.
%
%


\subsection{Geometric interpretation of $N_1$ and its presence in dense suspensions}
\label{111104_7Jun18}

As it is well known, the relative viscosity $\eta/\eta_0$ is a monotonically increasing function of the volume fraction $\phi$
(\figref{163218_21May18}).
It follows the functional form $\eta(\phi)/\eta_0 = c (\phi_{\mathrm{J}} - \phi)^{-\lambda}$ 
(solid lines in \figref{163218_21May18}), 
featuring a power-law divergence at a jamming point $\phi_{\mathrm{J}}$ 
that depends on the friction coefficient $\mu$ (vertical dashed lines in \figref{163218_21May18}).
As is expected in the presence of a similar divergence, we observe a growth of several orders of magnitude in the relative viscosity.


\begin{figure}
  \centering
  \includegraphics[width=1.\textwidth]{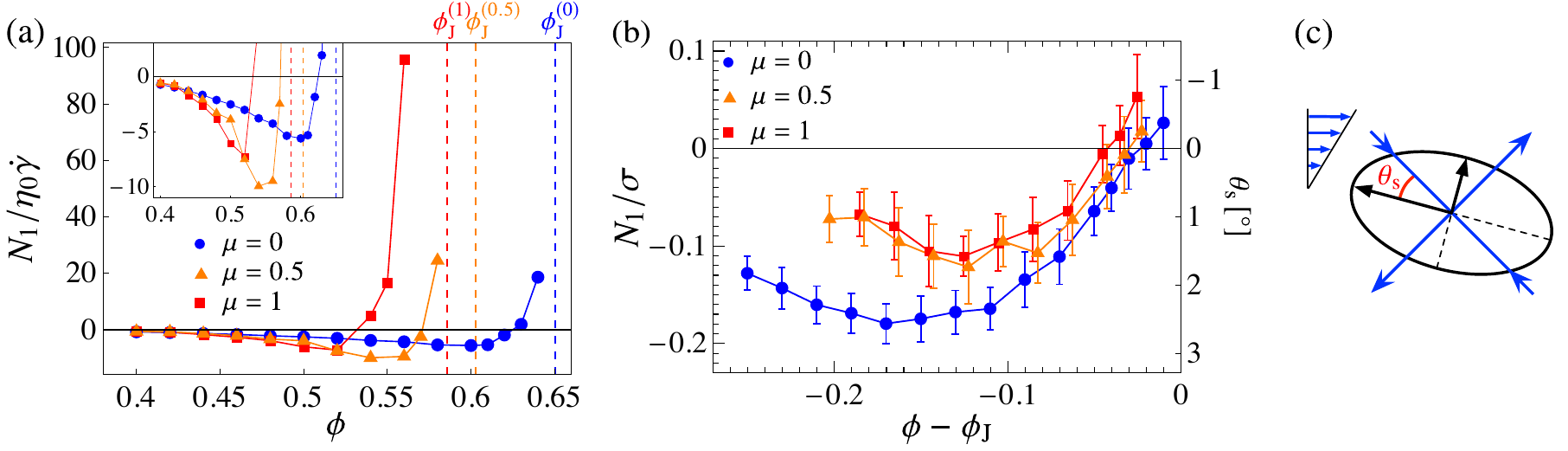}
  \caption{%
    (a) 
    The value of the normal stress difference $N_1$ divided 
    by the solvent stress $\eta_0 \dot{\gamma}$
    remains negative with an increasing intensity up to high volume fractions $\phi$, 
    where a sudden inversion towards large positive values is observed.
    The behavior is similar for different values of the friction coefficient $\mu$.
    Vertical dashed lines indicate the respective jamming points $\phi_{\mathrm{J}}^{(\mu)}$.
    (b) The ratio between $N_1$ and the shear stress $\sigma$ or the corresponding reorientation angle $\theta_{\mathrm{s}}$
    give a better idea of the mildness of the effect measured by $N_1$ over the whole range of explored volume fractions.
    (c) The reorientation angle $\theta_{\mathrm{s}}$ 
    is defined as the angle, in the flow plane, between the eigenvectors of the stress tensor $\bm{\sigma}$
    and the eigenvectors of $\Tensor{D}$ (oriented at $45^\circ$ from the flow direction).
  }
  \label{fig_n1_data}
\end{figure}


By contrast, the $\phi$-dependence of the first normal stress difference 
normalized by the solvent stress $\eta_0\dot{\gamma}$ is non-monotonic: 
it is negative and slowly decreasing for moderate volume fractions, 
reaches a minimum, and then rapidly increases to large positive values in the vicinity of the jamming point (\figref{fig_n1_data}\,(a)).
Nevertheless, a better appreciation of the role of $N_1$ and its rheological importance relative to that of the divergent viscosity 
comes from the analysis of the ratio $N_1/\sigma$ or the reorientation angle $\theta_{\mathrm{s}}$ (\figref{fig_n1_data}\,(b)).
In fact, a nonvanishing $N_1$ indicates that the eigenvectors of the stress in the flow plane are rotated with respect those of $\Tensor{D}$ by an angle
\begin{equation}
  \theta_{\mathrm{s}}
\equiv
  \tan^{-1}
  \Biggl(
    \frac{- N_1 / \sigma }{2 +
      \sqrt{4  + (N_1/\sigma)^2}}
    \Biggr),
\end{equation}
depicted in \figref{fig_n1_data}\,(c), which is determined 
only by the ratio $N_1/\sigma$~\citep{Giusteri_2018}.


In terms of these quantities, the measured values of $N_1$ are seen to correspond to a minor rheological feature---at least an order of magnitude smaller than the shear stress---that varies smoothly with the volume fraction.
Indeed, $N_1/\sigma$ is negative at lower volume fractions, decreases to
a minimum and then gradually increases towards zero, turning to positive values 
only in close proximity of the jamming point~(\figref{fig_n1_data}\,(b)).


\subsection{Synergy and competition between hydrodynamic and contact interactions}
\label{164152_5Jun18}

\begin{figure}
\centering
  \includegraphics[width=0.8\textwidth]{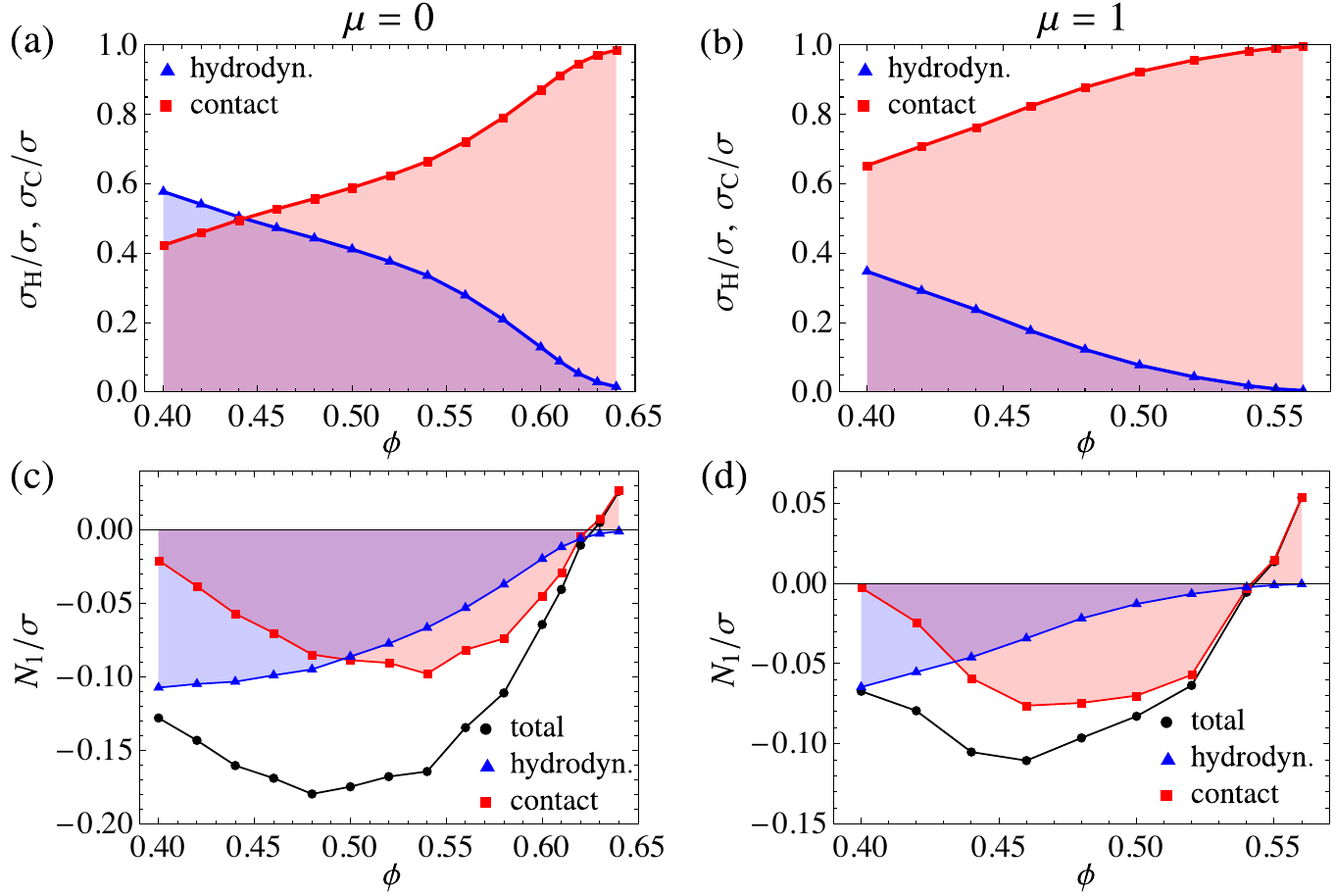}
  \caption{%
    (a,b)
    The contact contribution $\sigma_\mathrm{C}$ to the total shear stress $\sigma$ dominates 
    on the hydrodynamic one $\sigma_\mathrm{H}$ over almost the whole range of explored volume fractions, 
    both in the absence and presence of friction, even when hydrodynamic interactions determine a major fraction of $N_1$. 
    (c,d)
    Hydrodynamic and contact contributions to $N_1/\sigma$ for frictionless and frictional contacts are negative 
    over a wide range of volume fractions $\phi$.
    The hydrodynamic contribution decreases upon increasing $\phi$. The negative contact contribution becomes dominant but then vanishes, 
    before turning to positive values near the jamming point.
  }
  \label{102125_4Jun18}
\end{figure}

When discussing the $\phi$-dependence of the viscosity and, through $N_1/\sigma$, of the reorientation angle, 
it is instructive to breakdown the total values into hydrodynamic and contact contributions.
This is possible by taking advantage of the ``perfect knowledge'' offered by simulations, that is of course not available when treating experimental data.
As for the viscosity, upon increasing the volume fraction contact interactions become more and more likely to occur and progressively hinder hydrodynamic interactions.
We can pictorially say that the two effects ``fight for space'' and the shear stress $\sigma$ is mostly determined by contact interactions.
Moreover, as one would expect, the presence of friction enhances the preeminence of contacts (\figref{102125_4Jun18}\,(a,b)). 


The situation for the ratio $N_1/\sigma$ is quite different (\figref{102125_4Jun18} (c,d)).
At lower volume fractions the main contribution, with a negative sign, is given by hydrodynamic interactions, at intermediate volume fractions contacts begin to contribute, again with a negative sign, and the reorientation effect is strongest ($N_1/\sigma$ minimum) in a regime where the hydrodynamic contribution is still significant and the contact contribution is growing.
After this synergic regime, both the hydrodynamic and contact contributions approach zero and, close to jamming, only the contact contribution survives and switches to a positive sign.


It is thus clear that the change in the sign of $N_1$ near the jamming point
does not indicate the transition from a preeminence of contacts over hydrodynamic interactions, 
which mostly cooperate in building a microstructure that induces negative average contributions to $N_1$.


\subsection{From the microscopic force network to the macroscopic normal stress}

\begin{figure}
  \centering
  \includegraphics[width=1.\textwidth]{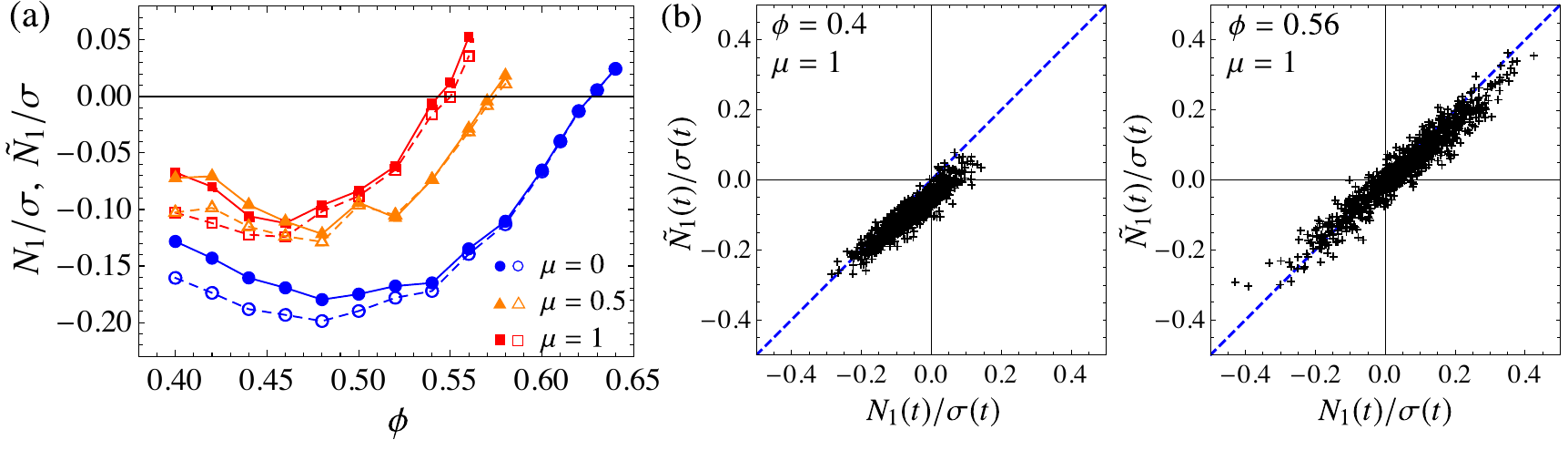}
  \caption{%
    (a) The reduced normal stress difference $\tilde{N}_1$ (open symbols)
    constructed by omitting tangential lubrication and contact forces
    can reproduce $N_1$ (solid symbols),
    especially at higher volume fractions.
    (b)~Even the instantaneous value $\tilde{N}_1(t)$
    linearly correlates to $N_1(t)$.
  }
\label{fig_n1_tilde}
\end{figure}

We still need to understand what determines the sign of $N_1/\sigma$.
To this end, 
we introduce a simplified quantity that retains the basic features of $N_1$.
Any force between particles $i$ and $j$
can be decomposed into two parts, normal and tangential forces,
by means of projections involving 
the normal vector $\bm{n}^{(ij)} \equiv (1/r_{ij})(\bm{r}^{(j)}-\bm{r}^{(i)})$.
Tangential contact forces play an essential role for the particle dynamics and rheology.
Indeed, the friction coefficient $\mu$ shifts the jamming point 
and also determines the behavior of $N_1$ as shown in \secref{111104_7Jun18}. 
However, 
we can verify that normal forces constitute the dominant part of the stress 
and especially the contributions of the tangential forces to 
$N_1$ are very minor.
Here, we introduce the reduced stress tensor including only normal forces,
\begin{equation}
  \tilde{\bm{\sigma}} \equiv 
V^{-1}\sum_{i > j} 
(\bm{r}^{(j)} - \bm{r}^{(i)}) \bar{\bm{F}}^{(ij)}, 
\end{equation}
where 
$\bar{\bm{F}}^{(ij)} \equiv 
(\bm{F}^{(ij)} \cdot \bm{n}^{(ji)}) \bm{n}^{(ji)}
= - \bar{F}_{ij} \bm{n}^{(ij)}$ 
is the normal force acting on particle $i$ from particle~$j$.
Here, positive (resp.\ negative) $\bar{F}_{ij} $ gives a repulsive (resp.\ attractive) force.
The reduced normal stress difference $\tilde{N}_1 $ is defined as
\begin{align}
  \tilde{N}_1 
  &\equiv \langle \tilde{\sigma}_{xx}- \tilde{\sigma}_{yy} \rangle \notag \\
  &=
  \Bigl\langle
    V^{-1} \sum_{i > j} r_{ij} \bar{F}_{ij}
    \Bigl[  \bigl(n_x^{(ij)}\bigr)^2 - \bigl(n_y^{(ij)}\bigr)^2 \Bigr] 
    \Bigr\rangle
  = 
  \Bigl\langle
    V^{-1} \sum_{i > j}
    \left( - r'_{ij} \bar{F}'_{ij} \sin 2\theta_{ij}\right)
    \Bigr\rangle
  = 
  \Bigl\langle
    \sum_{i > j} \tilde{N}_1^{ij}
    \Bigr\rangle,
\label{tildeN1}
\end{align}
where the local angle $\theta_{ij}$ is the one formed between 
the projection of the normal vector $\bm{n}^{(ij)}$ in the flow plane and the compression axis.
(Additionally, $r'_{ij} $ and $F'_{ij}$ are 
norms of the projected vectors on the plane.)
By analysing the data from our simulations,
we can confirm that 
not only $\tilde{N}_1$ is a good approximation of $N_1$ (\figref{fig_n1_tilde}\,(a)),
but even the instantaneous value $N_1(t)$ 
can be reasonably reproduced by the reduced $\tilde{N}_1(t)$,
especially at higher volume fractions where the contact forces are predominant,
as seen in \figref{fig_n1_tilde}\,(b).

\begin{figure}
  \centering
  \includegraphics[width=1.\textwidth]{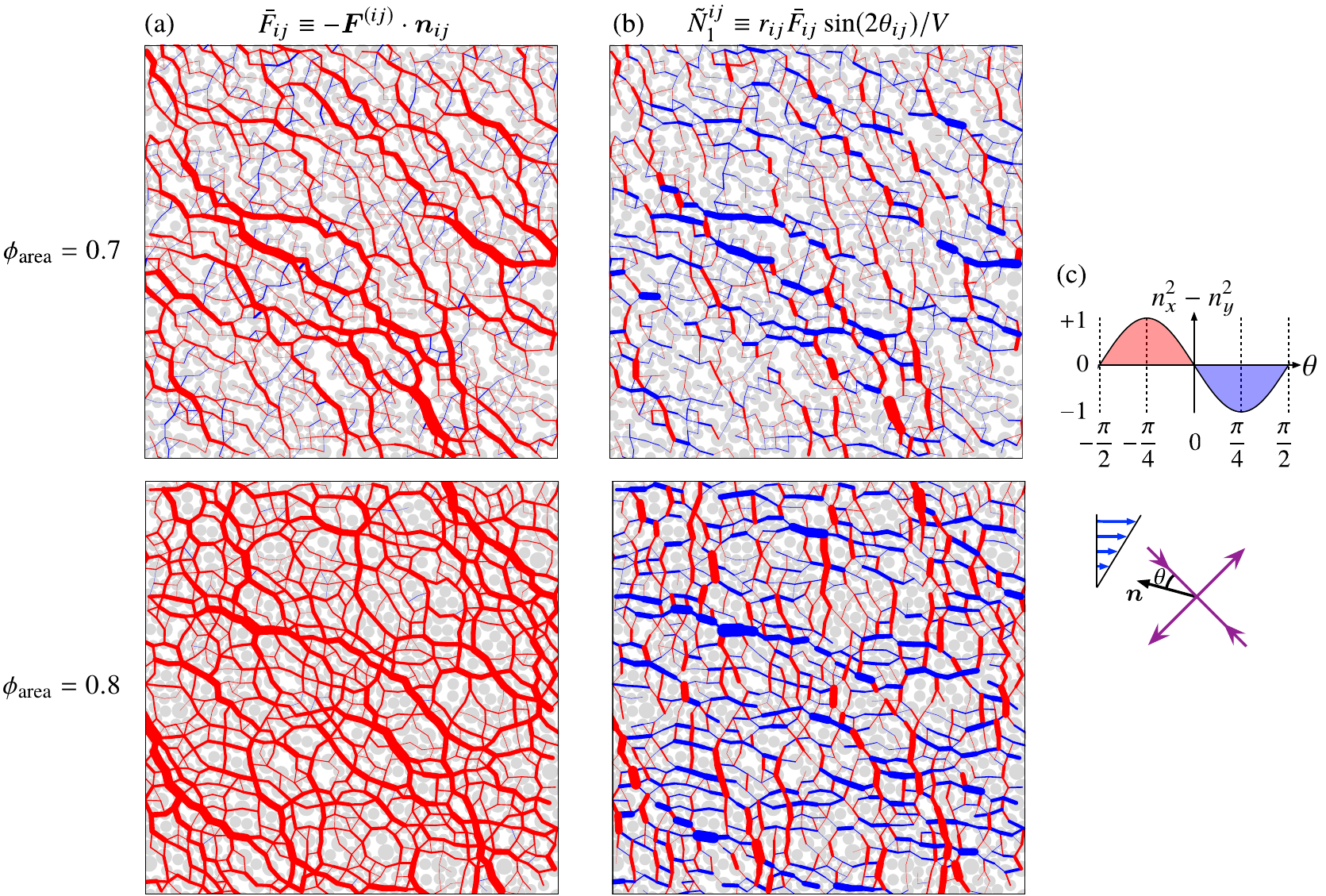}
  \caption{%
    (a) Snapshots of 2D monolayer simulations
    to show the force network of the pair-wise normal force $\bar{\bm{F}}^{(ij)}$.
    The thickness of the segments indicates the intensity of the force.
    Repulsive forces ($\bar{F}_{ij}>0$) are in red
    and attractive forces ($\bar{F}_{ij}<0$) in blue.
    (b) The pair-wise local contributions $\tilde{N}_1^{ij}$ to the first normal stress difference.
    The vertical repulsive forces contribute positively (red),
    while the horizontal ones negatively (blue).
    (c) Illustration of the factor $n_x^2 - n_y^2$, that determines the sign of $\tilde{N}_1^{ij}$, in terms of the local orientation angle $\theta$ formed by the normal force and the compression axis.
  }
\label{fig_n1_snapshot}
\end{figure}

To illustrate how the local contributors $\tilde{N}_{ij}$ relates to the force network,
we perform some 2D monolayer simulations,
which retain the qualitative features of the 3D simulations
while allowing for an easier visual perception, that can guide our understanding of the microscopic interactions.
\figref{fig_n1_snapshot}\,(a) presents typical snapshots of
the force network for a frictional system 
($\mu = 0.5$) at area fraction $\phi_{\mathrm{area}}=0.7$ (upper panel)
and $\phi_{\mathrm{area}}=0.8$ (lower panel).
Normal forces $\bar{\bm{F}}^{(ij)}$ between interacting pairs are drawn as segments.
The red and blue colors indicate repulsive and attractive forces.
The strength of the normal forces 
is visualized by varying the thickness of the segments.
We see that the most significant normal forces in these snapshots are repulsive (red).
%

The local contributors $\tilde{N}_1^{ij}$
to the reduced normal stress difference $\tilde{N}_1(t)$
are represented in a similar manner in \figref{fig_n1_snapshot}\,(b).
They contain the normal force $\bar{F}_{ij}$ as a factor (see \eqref{tildeN1}),
but it is the factor $n_x^2-n_y^2$ (a function of the local angle $\theta_{ij}$) that
decides the sign of the contribution to $\tilde{N}_1(t)$,
as shown in \figref{fig_n1_snapshot}\,(c).
(We observe that $\tilde{N}_1(t)$ is negative at $\phi_{\mathrm{area}}=0.7$
and positive at $\phi_{\mathrm{area}}=0.8$.)
When two particles align along the compression axis ($\theta_{ij}= 0$),
the normal force does not contribute to $\tilde{N}_1$ at all.
The instantaneous value of $\tilde N_1$ 
is given by the sum of many contributions, 
from all the particle pairs, with opposite sign. 
Indeed, in a typical snapshot, we can find strong local 
contributions of both positive (red) and negative (blue) sign.
%

\begin{figure}
\centering
\includegraphics[width=0.8\textwidth]{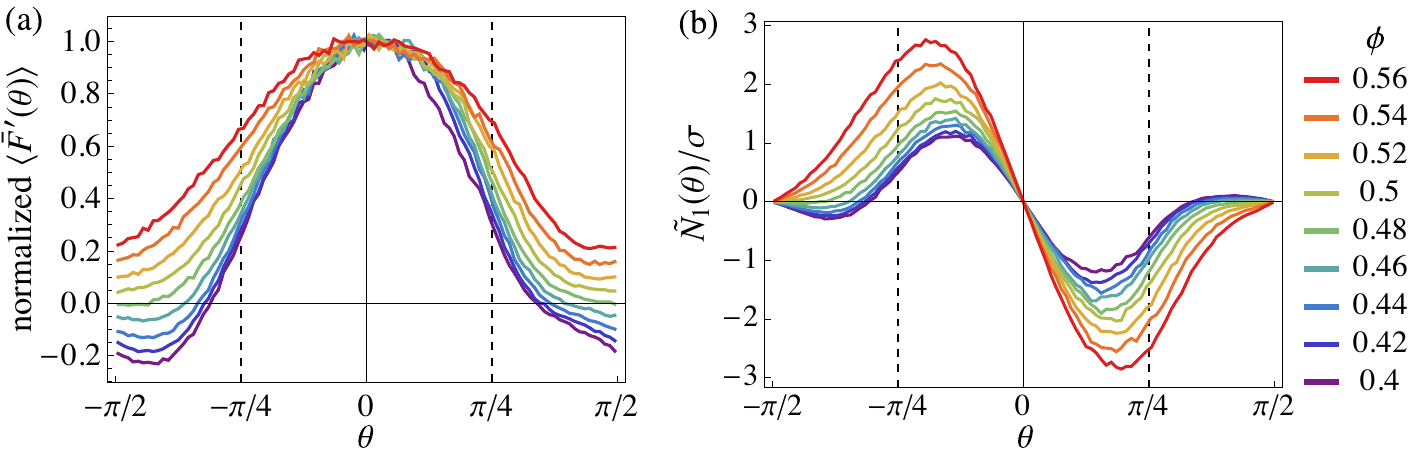}
\caption{%
  The significant normal forces are mostly repulsive and 
  the time-averaged values of $N_1/\sigma$ are the result of cancellations 
  among local contributions of opposite sign.
  This can be seen from the angular distributions of 
  (a) the projected normal force $\langle \bar{F}'(\theta) \rangle$
  on the flow plane
  and 
  (b) the ratio $ \tilde{N}_1 (\theta) /\sigma$,
  obtained from 3D simulations with friction coefficient $\mu  = 1$.
}
\label{160849_4Jun18}
\end{figure}

A more quantitative understanding of this observation can be reached 
by evaluating the time-averaged distribution of the normal forces 
and the normal stress components, $\bar{F}_{ij}$ and $\tilde{N}_1^{ij}$,
in terms of the local angle~$\theta$ 
for 3D simulations.
\figref{160849_4Jun18}\,(a) 
shows the distribution $\langle \bar{F}(\theta) \rangle$ of the normal forces divided by the maximum value $\bar{F}_{\mathrm{max}}$.
The peaks of the force distributions are found around the compression axis ($\theta = 0$).
This confirms that,
though we can find some attractive forces (negative values)
along the extension axis at lower values of $\phi$,
the significant normal forces are mostly repulsive.

The angular distributions $ \tilde{N}_1(\theta) /\sigma$ 
in \figref{160849_4Jun18}\,(b)
show very prominent positive and negative peaks.
Nevertheless,
the global value of $\tilde{N}_1/\sigma$, 
obtained
by integrating $\tilde{N}_1(\theta)/\sigma$ 
over the entire range of directions,
is somehow small with large fluctuations over time.
The fact that the global values are the result of a cancellation 
between large contributions of opposite signs
makes the sign of its fluctuating instantaneous values uncertain.


We can conclude that a global nonvanishing value of $N_1$ is generated by a small imbalance of positive and negative contributions in the angular distributions presented above.
This is associated with a mild preference of the dominant branches of the force network to be aligned along a direction different from the compression axis defined by the shear flow.


\subsection{Anisotropy due to the planarity of the flow}

\begin{figure}
  \centering
  \includegraphics[width=0.8\textwidth]{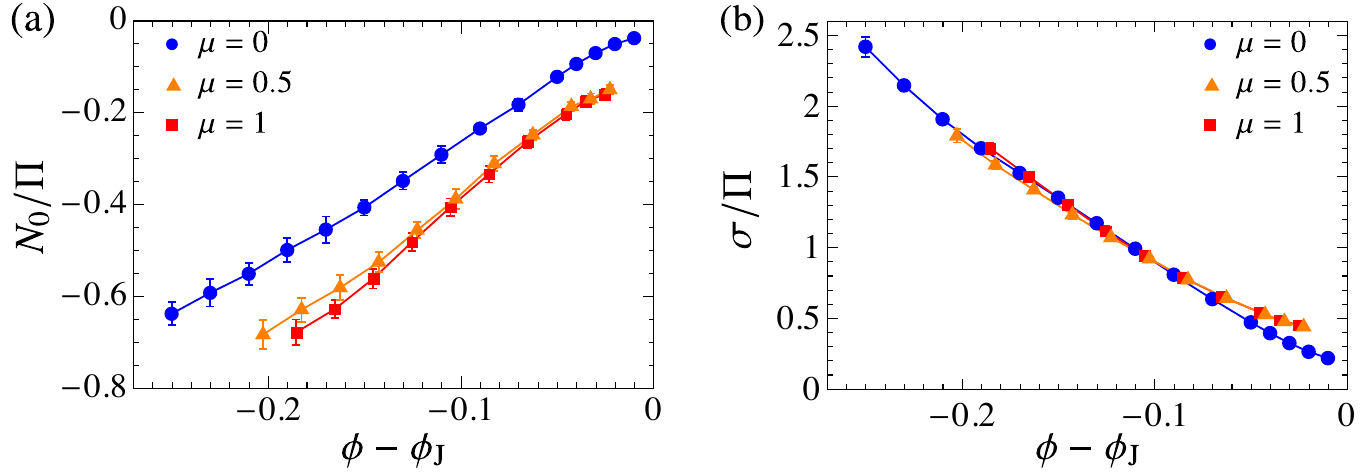}
  \caption{%
    (a) The absolute value of the ratio $N_0/\Pi$, 
    a measure of the anisotropy of the stress in the vorticity direction, 
    decreases upon increasing the volume fraction $\phi$
    since the force network becomes more and more isotropic as the jamming point is approached.
    For frictional systems, a residual anisotropy is observed.
    (b) The ratio of $\sigma$ to $\Pi$,
    a measure of the anisotropy in the flow plane,
    also decreases with $\phi$.
    The data for different values of the friction coefficient $\mu$
    display a nice collapse as a function of $\phi-\phi_{\mathrm{J}}$,
    except for the points close to the jamming condition.
  }
  \label{fig:lambda0}
\end{figure}

As mentioned in the introductory section, the stress tensor 
in steady simple shear is characterized by three degrees of freedom.
So far, we discussed the $\phi$-dependence of the viscosity $\eta = \sigma/\dot{\gamma}$ 
and the first normal stress difference $N_1$.
The third degree of freedom is normally associated with the second normal stress difference $N_2$ but, 
as shown by \citet{Giusteri_2018}, this quantity is not fully independent of $N_1$, 
since it is a combined measure of the misalignment between $\CauchyStress$ 
and $\Tensor{D}$ (already captured by $N_1$) and of a second effect.
The latter corresponds to a stress contribution which is isotropic in the flow plane (i.e., the $x$-$y$ plane)
but globally anisotropic, when the vorticity direction is taken into account.
In a simple shear flow, the nonvanishing vorticity is everywhere orthogonal to the flow plane and the invariance under any translation along this direction characterises the planarity of such flows.
A good measure of the third independent degree of freedom is given by
\begin{equation}
  N_0 \equiv
\left\langle\frac{\sigma_{xx}+\sigma_{yy}}{2}
- \sigma_{zz} \right\rangle
  =  N_2+\frac{N_1}{2}.
  \label{eq:N_0}
\end{equation}
$N_0$ is normalized by the isotropic pressure $\Pi \equiv -(1/3)\langle \tr \CauchyStress \rangle$,
so that $N_0/\Pi$ can be understood 
by considering that it reflects an anisotropy of the normal stress
(or ``pressure'') originated from the planarity of the flow.
The negative values of $N_0/\Pi$ in \figref{fig:lambda0}(a)
indicate that 
the flow generates more ``pressure'' in the flow plane
than in the vorticity direction.
This anisotropy monotonically decreases as the system becomes denser and denser,
and the viscosity higher and higher (see \figref{163218_21May18}).
In the limit where the viscosity of frictionless systems ($\mu = 0$) diverges,
the anisotropy looks to vanish.
This is consistent with the idea that flow-induced microstructures are not relevant 
to frictionless jamming, being this dominated by geometric effects.
On the other hand,
a residual stress anisotropy survives 
in the limit of jamming with friction $\mu>0$.
Indeed, it has been suggested that flow-induced microstructures may contribute
to the jamming of frictional systems~\citep{Cates_1998a,Bi_2011}. 
A similar observation was also reported as ``absence of dilatancy''
in the quasi-static limit of frictionless granular flows~\citep{Peyneau_2008}.


To complete our discussion of the stress anisotropy, 
we now consider the anisotropy generated in the flow plane by the shear flow itself.
This is of course a dominant effect in shear rheology and it can be measured by the ratio of 
the shear stress $\sigma = \eta \dot{\gamma}$ to $\Pi$, 
a quantity that defines the macroscopic friction coefficient in the context of granular flows~\citep{Boyer_2011}.
The ratio $\sigma/\Pi$ displays a decreasing trend upon increasing the volume fraction, 
but a finite anisotropy seems to survive even in the proximity of jamming (\figref{fig:lambda0}(b)).
Notably, for a range of volume fractions that are not too close to the jamming point, the data for different friction coefficients collapse on the same curve.
This points to a geometric origin of this anisotropy in dense suspensions, that makes it insensitive to friction, which is in turn essential in determining dynamical properties of the system such as the intensity of the stress response. 
Nevertheless, close to jamming we observe a measurable difference 
in the residual anisotropy for frictional and frictionless systems.
Such difference seems equivalent
to what is observed in the quasi-static limit of granular dynamics, that is $\sigma/\Pi  \sim 0.1$ for $\mu = 0$~\citep{Peyneau_2008}
and
$\sigma/\Pi \sim 0.35$ for $\mu > 0.4$~\citep{Singh_2013,Azema_2014}.

\subsection{The role of elastic effects near jamming}
\label{123947_14Jun18}

In view of the link we established between the microscopic properties 
of the force network and the macroscopic value of $N_1$, 
the presence of a nonvanishing $N_1$ near jamming appears to be puzzling, 
irrespectively of its sign.
In the limit of jamming with hard spheres,
we expect that effects of the rotational component of the flow 
are negligible compared to those of 
the extensional component~\citep{Seto_2017}.
If so, the force network would become statistically symmetric 
across the compression axis in the flow plane, entailing a vanishing $N_1$.
Nevertheless, our data indicate the presence of a symmetry breaking measured by nonzero values of $N_1$.
We thus need to identify some factor that has been ignored in our previous arguments.


In analogy with what happens in viscoelastic fluids, the symmetry could be broken by the vorticity if some elastic links were actually convected by the flow.
While there is no such link in a hard-sphere suspension, 
the contact model employed in our simulation effectively approximates 
hard spheres with high-stiffness elastic ones.
We then argue that the observed positive values of $N_1$ are 
due to a failure of the simulation strategy in resolving contacts in a sufficiently rapid way.
Particles that overlap significantly for some time produce normal forces with directions that are rotated towards the gradient direction, inducing an average reorientation of the stress eigenvectors.

\begin{figure}
\centering
  \includegraphics[width=\textwidth]{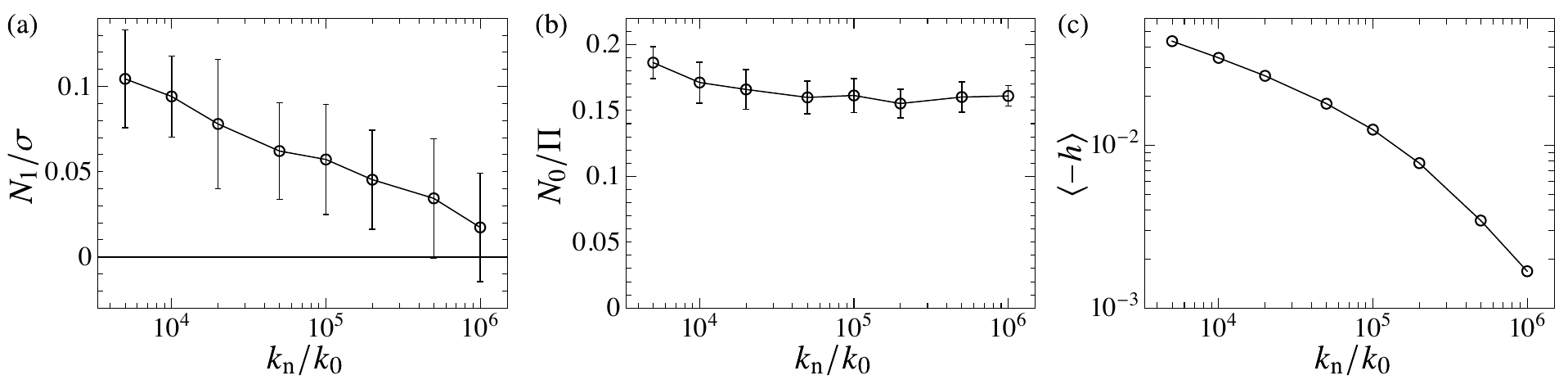}
  \caption{%
    (a) The positive value of $N_1/\sigma$, 
    observed at the volume fraction $\phi = 0.56$ with friction coefficient $\mu = 1$,
    decreases if we increase the elastic constant $k_{\mathrm{n}}$
    employed in the contact model to approximate the hard-sphere interaction.
    This indicates such elastic interactions as the origin of the positive $N_1$ near jamming.
    (b) The observed value of $N_0/\Pi$ is not significantly affected by changes in $k_{\mathrm{n}}$ over the explored range.
    (c) 
    The dependence of the average maximum overlap $\langle -h \rangle$ on $k_{\mathrm{n}}$ 
    indicates that our simulation is  gradually approaching the hard-sphere limit  
    as the elastic constant $k_{\mathrm{n}}$ increases.
  }
  \label{125445_7Jun18}
\end{figure}

To confirm this interpretation, we analysed the dependence of $N_1/\sigma$ 
on the effective normal stiffness $k_{\mathrm{n}}$ of the frictional contact model ($\mu=1$) 
by running 20 independent 3D simulations for each value of $k_{\mathrm{n}}$ 
at the volume fraction $\phi = 0.56$, 
which is close to jamming and gave a positive value of $N_1$ in the data reported above.
(The maximum displacement 
is set to a common constant value $d_{\mathrm{max}}= 5 \times 10^{-4}a$ for $k_{\mathrm{n}} \leq  10^{5}k_0$ and,
to ensure the reliability of the simulations, reduced 
in inverse proportion to $k_{\mathrm{n}}$ for stiffer particles with $k_{\mathrm{n}} >  10^{5}k_0$.)
Even though we cannot test extremely large values of $k_{\mathrm{n}}$ 
(the time steps would get extremely short and the simulation time diverge) 
we found a clear trend of $N_1/\sigma$ decreasing 
as $k_{\mathrm{n}}$ increases (\figref{125445_7Jun18}\,(a)).
We may infer that, in the hard sphere limit, 
$N_1 /\sigma$ is negative below jamming and 
asymptotically approaches zero at the jamming point.
In the same conditions, the value of $N_0/\Pi$ 
is not affected by the tested increase in $k_{\mathrm{n}}$ (\figref{125445_7Jun18}\,(b)).
This means that the finite stress anisotropy observed near the frictional jamming and 
due to the planarity of shear flows is not an artifact of the simulation but a genuine phenomenon.
%


Based on the awareness we gained from the computational results for stiffer particles, 
we can make an instructive comparison with experimental studies. 
We take the work by \citet{Royer_2016} as an example.
In these experiments, a suspension of silica particles with $2a=\SI{1.54}{\micro\metre}$
is fully thickened under a shear stress of \SI{6000}{\pascal}
and exhibits a positive $N_1$ for volume fractions above $\phi = 0.54$.
The typical contact deformation in such situation
can be estimated as $10^{-4} a$, 
by using the Hertzian contact model $F =(4/3)E^{\ast} a^{1/2} h^{3/2}$
with an effective elastic modulus $E^{\ast} =  E/2(1-\nu^2)$ 
given by assuming the Young modulus $E=\SI{70}{\giga\pascal} $ 
and Poisson ratio $\nu = 0.17$ of silica particles.
Such average overlaps can be realized
with $k_{\mathrm{n}}/k_0 \approx 10^{7}$ in our simulation,
if the results of \figref{125445_7Jun18}\,(c) are simply extrapolated.
For such stiff particles, the computational value of
$N_1 /\sigma$ is expected to be vanishing or slightly negative
from extrapolating the data presented in \figref{125445_7Jun18}\,(a).
Nevertheless, the data from \citet{Royer_2016}, red circles in \figref{172630_30Jul18}, show definitely positive values of $N_1 /\sigma$.
The discrepancy between simulations and experiments can be explained either 
with the presence of interactions absent from our model, 
if it concerns bulk rheology, or as a signature of boundary effects due to the presence of walls in standard rheometers.
Indeed, \citet{Gallier_2016} numerically investigated wall effects on $N_1$
and found that $N_1 /\sigma$ 
can be largely positive due to wall-induced ordering.


\section{Conclusions}

We investigated, by means of particle dynamics simulations, 
the presence and the microscopic origin of normal stress differences
in dense suspensions under simple shear flows in the high-P\'eclet-number limit.
By interpreting the first normal stress difference $N_1$ as a measure of the misalignment between the stress $\CauchyStress$ and the symmetric part of the velocity gradient $\Tensor{D}$, we have shown that it represents a minor effect in comparison to the increment in the viscous response due to the interactions among the dispersed particles.
Importantly, we provided evidence that the sign of $N_1$ cannot be used to discriminate whether hydrodynamic or contact interactions are dominant.
In fact, in the dense regime, hydrodynamic and contact interactions always cooperate to give negative contributions to $N_1$.
From our analysis, it appears how the properties of the force network 
generated under shear are key to understand the rheology of the system.
Indeed, the observed misalignment is so mild because it originates 
from a small imbalance of intense but competing local contributions, 
that cancel each other in the macroscopic average.

\begin{figure}
\centering
  \includegraphics[width=0.7\textwidth]{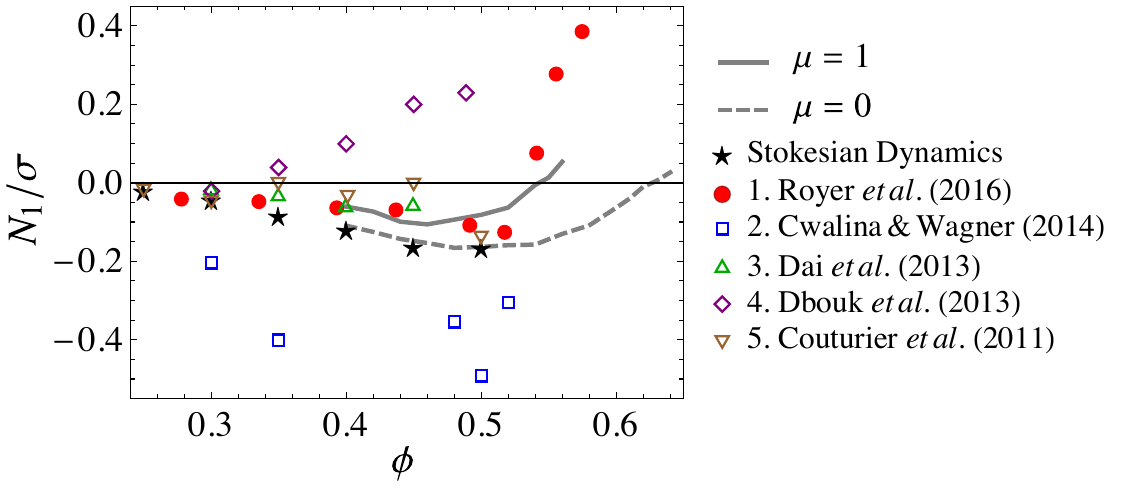}
  \caption{%
    Comparison of our simulation results for the cases $\mu = 1$ (solid gray line)
    and $\mu = 0 $ (dashed gray line)
    with some reported experiments from the literature.
    Our frictionless simulation 
    agrees with the Stokesian Dynamics by \citet{Sierou_2002}, 
    indicating that the lubrication approximation is justified at such high volume fractions.
    The frictional simulation near jamming shows a value of $N_1/\sigma$ much closer to zero than the reported experimental data,
    suggesting that some effects are being neglected in the computational model.
    The material and diameter of the particles used in the experiments 
    are:
    1. Silica, $\SI{1.54}{\micro\meter}$;
    2. Silica, $\SI{0.52}{\micro\meter}$;
    3. Polystyrene, $\SI{40}{\micro\meter}$;
    4. Polystyrene, $\SI{40}{\micro\meter}$;
    5. Polystyrene, $\SI{140}{\micro\meter}$.
  }
  \label{172630_30Jul18}
\end{figure}

Moreover, microscopic arguments allow to understand the meaning of positive values of $N_1$. 
For hard-sphere suspensions close to the jamming condition the force network 
becomes symmetric across the compression axis in the flow plane, implying a vanishing $N_1$.
This is a clear discrepancy with the positive values of $N_1$ found in some experiments.
We argue that those values ought to be traced back to effects that cannot be accounted for with simple hard-sphere models.
In fact, we show that the positive values obtained in our simulations are due to the presence of elastic interactions employed to regularize the the (numerically stiff) hard-sphere constraint.
A possible explanation of the experimental results could be sought in the presence of 
persistent elastic interactions between particle pairs that are advected by the flow.
However, as shown in \figref{172630_30Jul18},
definitely positive values of $N_1/\sigma$ near jamming
are found in experiments with particles that are much stiffer than in our simulation,
such as those by \citet{Royer_2016}.
This may suggest boundary effects due to the presence of walls in experiments, rather than bulk rheological properties, as an alternate explanation
for the observed positive values of $N_1/\sigma$.

In place of the second normal stress difference $N_2$, 
we studied the quantity $N_0 \equiv  N_2 + N_1/2$.
This measures an effect genuinely independent of the misalignment measured by $N_1$, namely a stress contribution isotropic in the flow plane but globally anisotropic.
It reflects an anisotropy in the force network originated from the planarity of the flow.
The force network tends to be isotropic near jamming for frictionless contacts, and $N_0$ vanishes accordingly, 
but some residual anisotropy is observed near jamming for the case of frictional contacts.


In this work, we restricted our attention to the high-P\'eclet-number limit,
where only hydrodynamic and contact forces determine the particle dynamics.
Under lower stresses,
some other force, such as Brownian forces or short-range repulsive forces,
tends to prevent contact and to maintain a lubrication layer between particles.
Such lubricated contacts are more similar to those obtained in the frictionless system.
Under higher stresses, the effect of the additional force declines and frictional contacts appear.
As a consequence of this mechanism,
the shear thickening of dense suspensions, a rate-dependent feature, can be reproduced
by interpolating between two points 
on the rate-independent frictionless and frictional rheology curves in \figref{163218_21May18}
with a function of a state parameter that controls the effective jamming point~\citep{Wyart_2014,Singh_2018}.
Analogously, a possible way to predict the rate dependence of $N_1/\sigma$ would involve 
interpolating the reported results for frictionless and frictional systems.
Nevertheless, the non-monotonic behaviour of $N_1$ makes it harder to find a suitable interpolating function.

\section*{Acknowledgments}

This study was supported by Japan Society for the Promotion of Science (JSPS) 
KAKENHI Grants No.~JP17K05618
and New Energy and Industrial Technology Development Organization of Japan (NEDO) Grant No.\ P16010.
The research of R.S. was also supported in part by the National Science Foundation under Grant No. NSF PHY-1748958. 
The authors would like to thank the participants of the KITP program ``Physics of Dense Suspensions'' 
for the many fruitful discussions and A.~Singh, L.~Hsiao, M.~Denn, and J. Morris for providing useful comments on the manuscript.
%


\end{document}